\documentclass[11pt, a4paper]{article}
\usepackage{moriond,epsfig}

\def\be{\begin{equation}}
\def\ee{\end{equation}}
\def\bea{\begin{eqnarray}}
\def\eea{\end{eqnarray}}

\def\kms{{km\thinspace s$^{-1}$}}
\def\13co{$^{13}$CO}
\def\h2{H$_2$}

\def\be{{$b$}\ }
\def\deg{\ifmmode^\circ\else$^\circ$\fi}
\def\dege{{\ifmmode^\circ\else$^\circ$\fi}\ }
\def\solar{\ifmmode _{\mathord\odot}\else$_{\mathord\odot}$\fi}
\def\sun{$_\odot$}

\def\asece{$^{\prime \prime}$\ }

\def\amine{$^{\prime}$\ }

\def\hper{\ifmmode \rlap.^{h} \else $\rlap{.}^h$\fi}
\def\mper{\ifmmode \rlap.^{m} \else $\rlap{.}^m$\fi}
\def\sper{\ifmmode \rlap.^{s} \else $\rlap{.}^s$ \fi}
\def\degper{\ifmmode \rlap.^{\circ} \else $\rlap{.}^{\circ} $\fi}
\def\arcmper{\ifmmode \rlap.{' } \else $\rlap{.}' $\fi}
\def\arcsper{\ifmmode \rlap.{'' } \else $\rlap{.}'' $\fi}

\def\c2{cm$^{-2}$}
\def\cm2{cm$^{-2}$}

\def\c2{cm$^{-2}$}
\def\cm2{cm$^{-2}$}
\def\cm2e{{cm$^{-2}$}\ }

\def\apjs{{ApJS},\ }      
\def\apjl{{ApJL},\ }           
\def\apj{{ApJ},\ }       
\def\mnras{{MNRAS},\ }               
\def\aasup{{A\&AS},\ }       
\def\araa{{ARA\&A},\ } 
\def\pasj{{PASJ},\ }              

\begin{document}
\vspace*{4cm}
\title{GLOBAL STAR FORMATION FROM ${\bf z = 5 \times 10^{-8}}$ to ${\bf z =
20}$}


\author{LEO BLITZ}
\address{Astronomy Department, University of California\\
Berkeley, California  94720, USA}

\maketitle\abstracts{Starting with the assertion that the 
problem of isolated, single star formation is essentially solved, 
this paper examines some of the missing steps needed to go from there 
to understanding the star formation history of the Universe. Along the
way, some results on the formation of star
clusters in the Milky Way and the properties of GMCs in nearby galaxies 
are briefly examined. 
} 

\noindent
{\small¥{\it Keywords}: Stars: formation; ISM: general; Galaxies:
evolution}

\section{Introduction} 
The field of star formation exploded with the advent of
millimeter-wave and infrared detectors in the 1970s.  Prior to that it was a
field with a few lonely but brilliant workers such as George Herbig and Adriaan
Blaauw who managed to identify young stars and regions of star formation from
their optically determined properties alone.  Both realized that the regions of
most recent star formation were always associated with dark dust clouds, and
understood that the earliest stages of star formation would only be probed by
penetrating the veil of dust obscuration. Since that time, the field of star
formation has expanded to include not just the nearest accessible regions, but
the farthest reaches of the Universe as well.  Using what we've learned about
local star formation, reasonable speculations and simulations have now been
attempted to guess at what the first stars in the Universe might have been like
(Abel, Bryan \& Norman 2002)~\cite{ref:abel02}.

The field of star formation remains a rich area of research with many
unsolved problems and thus continues to attract a coterie of young
inventive scientists.  In this article I give a personal view of where
some parts of the field are headed, especially those areas that touch
on star formation in galaxies.  Clearly, in such a short 
space, I can cover only a few topics, and even those, rather cursorily.     

\section{What Do We Really Want to Know?}

\begin{figure}[h!]
\begin{center}
\vskip 1.0cm
\psfig{figure=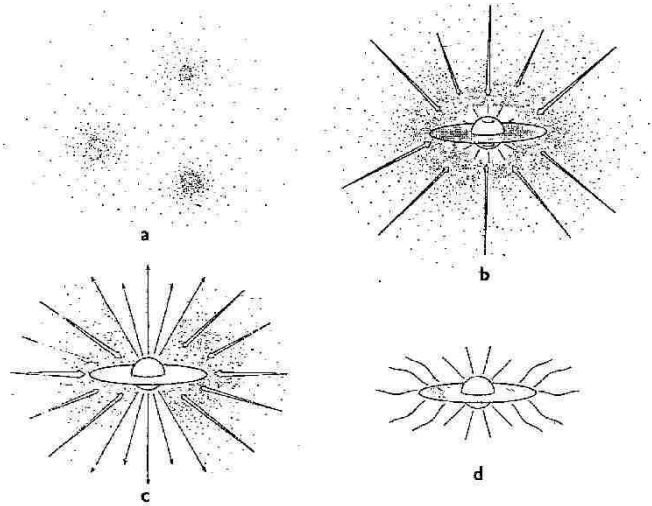,height=7cm}
\caption{Figure taken from the 1987 review of low-mass star formation
by Shu, Adams, and Lizano, giving the four principle stages of star
formation.  Although there has been some refinement of these stages,
there is general agreement among astronomers that star and solar
system formation occur according to this picture, which allows
detailed quantitative estimates to be made to compare with
observations.
\label{fig:shufig}}
\end{center}
\end{figure}

I begin by making the outrageous claim that the problem of low-mass,
single star formation is essentially solved, due in large part to the
work of Frank Shu, Richard Larson and a number of others.  This is not
to say that there aren't still questions that are worth asking, but
that the most interesting questions are more in the realm of planet 
formation than star formation. The remaining 
star formation issues are questions 
more of detail, rather than questions of a fundamental nature about 
how stars form. I illustrate this point with reference to what I
believe is the most well known image in the scientific literature
on the subject of star formation, which is shown here as Figure
\ref{fig:shufig} (Shu, Adams \& Lizano 1987)~\cite{ref:shu87}. 
(Jeff Hester's beautiful image of the elephant trunk
structures in the Eagle Nebula is more well known, but is reproduced
primarily in the popular press). 

This sketch represents the four stages of star formation which are now
generally accepted as how low-mass, single stars form.  Stars
begin as gravitationally unstable condensations in cold, dense
molecular clouds, forming a prestellar core observable in the near
infrared as material continues to rain in on it.  The higher angular
momentum material forms a disk, and the system develops a
bipolar outflow and jet which removes the angular momentum from the
system, while initially disrupting and clearing out the infalling
material.  The star becomes visible as a T-Tauri star, and the disk
ultimately becomes the raw material from which planets form.  Magnetic
fields play a central role in the dynamics, and add computational
complexity, but almost surely determine the onset of collapse
and the bipolar outflows.  Diverse observations have a good
theoretical underpinning, and little work is now done without either
explicitly or implicitly invoking this picture.

\begin{figure}[h!]
\begin{center}
\vskip 1.0cm
\psfig{figure=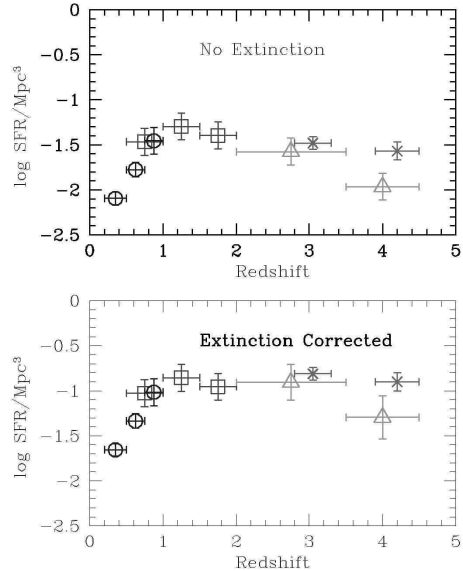,height=8cm}
\caption{The star formation rate in the universe as a function of
redshift.  The units on the ordinate are M\sun~Mpc$^{-3}$~yr$^{-1}$
based on the work of  (Steidel et al. 1999).
\label{fig:madau}}
\end{center}
\end{figure}

On the other end of the distance scale, deep observations with the
HST, Keck and SCUBA, have made it possible to determine the star
formation history of the Universe, which is shown in Figure
\ref{fig:madau} (Steidel et al. 1999)~\cite{ref:steidel99}.  This
plot, widely known as the Madau plot (Madau et al.
1996)~\cite{ref:madau96}, has been modified by others (e.g.
Rowan-Robinson 2002~\cite{ref:rr02}), but its main features are well
established: at early times there is a constant, or nearly constant star
formation rate per comoving Mpc until about $z$ = 2, at which time
the star formation rate steadily falls by about an order of magnitude
until the present epoch.  One of the great challenges is to apply what
we know about the details of star formation in the nearest regions to
fill in the missing pieces needed to obtain Figure \ref{fig:madau}.
The goal is to obtain not only the correct shape, but the correct
amplitude of the Madau function.

\subsection{Some of the Missing Pieces} 

\subsubsection{High Mass Star Formation}

The first problem is that the star formation rate in the
Universe is determined from the light of the most massive stars, but most
of what we know about star formation applies only to low mass stars.
Getting an evolutionary picture of high mass star formation remains
difficult observationally because of the rapid destruction of their
surroundings by high mass protostars. Without a good set of
observations it is difficult to make progress in the theory.  For
example, only a small number of candidate high mass prestellar cores
have been identified, and little has been written about the
relationship of these cores to their surrounding molecular clouds.
Nevertheless, considerable progress should be possible in the near
term from observations with a new generation of millimeter-wave
interferometers: CARMA and ALMA, the Spitzer Space Telescope, and
SOFIA. 

\subsubsection{The Formation of Stars in Clusters}

If the problem of how individual low-mass stars form is essentially
solved, and if high-mass star formation is next, we only get to
the first rung of the ladder that ends at the Madau plot. Stars do not
typically form as isolated objects, but rather in clusters, and little
is known about clustered star formation.  For example, do massive
clumps form massive star clusters?  What determines the star formation
efficiency of a particular cluster forming clump? Are the stars that
form in a cluster different from those that form in looser aggregates?
Are the prestellar cores that form star clusters in the gas clumps 
the same as those identified with single star formation?

The study of clustered star formation is in such a primitive state
that even some of the most basic questions have not yet been
addressed. For example, it would seem that high mass stars are the
last to form in clusters (lest they dissociate the gas from which the
accompanying low mass stars form), and they appear to form in the
cluster centers.  But how can this be the case since high mass
protostars should have the shortest dynamical times, and if formed in the
centers of the clumps, should form first since the density of the
gas is highest there.  Although there have been several guesses at a  
solution (e.g. Stahler, Palla \& Ho
2000~\cite{ref:stahler2000}; Bonnell, Bate \& Zinnecker
1998~\cite{ref:bonnell1998}), no explanation seems
compelling yet.

\subsubsection{Universality of the IMF}

The calibration of the vertical scale in 
Figure \ref{fig:madau} assumes that the IMF is invariant at all epochs
and in all galaxies. But how universal is the IMF?  To predict how it
might or might not vary in other galaxies, and at other epochs, we
need to know what physical or stochastic processes determine it. Very
little is known about how the initial mass function is produced,
though new work by Shu, Li, \& Allen (2004)~\cite{ref:shu04} promises
some progress on that subject.   

\subsubsection{The Formation of Stars in Galaxies}

It may be a long time before it is possible to understand enough about
the details of star formation to predict how star
formation proceeds on galactic scales in GMCs.  Nevertheless, it may
be possible to circumvent this issue by learning how GMCs form and
then determining how star formation proceeds {\it on average} in these 
GMCs.  With this approach one would need to know how the
physical conditions in GMCs differ in various galaxy types, in
different locations within a galaxy, and with changes in metallicity.
While this may seem like a daunting task, improvements in
instrumentation now make it possible to survey entire galaxies at high
enough resolution to make significant progress.  Some early results
are discussed below.

The last step in getting to the Madau plot is then extrapolating what
we know about global star formation in normal galaxies to the star
formation in starbursts and AGN.  In other words, why do particular
galaxies become starbursts, and how much star formation comes from
particular galaxy or merger?  This step is important because a
significant fraction of the light of galaxies comes from starbursts,
and the fraction of starbursts seems to change with $z$.

\subsubsection{Initial conditions, Initial Conditions, Initial
conditions}

What ties all of these points together is that to make the step from
single star formation to the Madau plot, it is necessary not only to
learn about the physical processes involved, which requires a
combination of theory and observation, but to understand what the
initial conditions are that give rise to variations in each step. For
example, even if the process of isolated single star formation is
essentially solved, we really have no idea how the intial conditions,
the star forming cores, are produced. Furthermore, we
don't know whether the IMF reflects the mass spectrum of
prestellar cores as suggested by Motte, Andre \& Neri
(1998)~\cite{ref:motte98} and Testi \&
Sargent(1998)~\cite{ref:testi98}, 
or the process of star formation itself (Shu, Li, \&
Allen 2004~\cite{ref:shu04}).  The beautiful work by Alves, Lada \& Lada
(2001)~\cite{ref:alves01} suggests that the initial configuration for star formation may
be better represented by a Bonner-Ebert sphere rather than a singular
isothermal sphere (Shu 1977)~\cite{ref:shu77}. 
Does this make a significant difference
in the star that is produced?  What are the initial conditions in a
GMC that produce the difference between relatively isolated star
formation (as in Taurus) and clustered star formation (as in
Orion)?  What are the initial conditions that give rise to the
number and distribution of GMCs in a galaxy?  

My own view is that we cannot know too much about typical initial
conditions and how they vary. Therefore, there cannot be too
much emphasis on trying to determine what the intial conditions are
for forming individual low mass and high mass stars, for forming stars
in clusters (why, for example, do some become globular clusters?), and 
for forming GMCs in the disks and centers of normal and starburst
galaxies. 

\section{A Few Relevant Results}
 
\subsection{Beyond Single Star Formation}

\begin{figure}[!h]
\begin{center}
\vskip 1.0cm
\psfig{figure=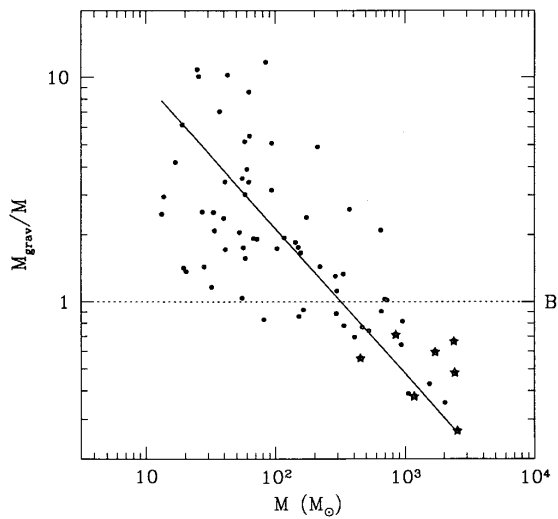,height=7cm}
\caption{Plot of the ratio of gravitational binding energy to kinetic 
energy of the clumps in the Rosette Molecular Cloud from Williams, et
al.(1995).  The horizontal line represents a
clump that is marginally self gravitating; clumps that are not
self-gravitating are presumably pressure bound.  The clumps containing
IRAS sources are shown with stars.  These are largely coincident with
the embedded dlusters found in the infrared by Phelps \& Lada 
(1997). 
\label{fig:ros.ps}}
\end{center}
\end{figure}

One of the first attemps to study to study the formation of star
clusters observationally 
was done by Elizabeth Lada (Lada 1992)~\cite{ref:lada92}, who made the
first survey of dense gas in an entire GMC (Orion B) using the 
molecular tracer CS.  
She found that the embedded stars are found
primarily in clusters, and that the clusters form
only in the densest condensations: those identified by their CS
emission.  Subsequently, Phelps \& Lada (1997)~\cite{ref:phelps97}
made another advance with their near IR imaging of some of the $^{13}$CO
clumps in the Rosette Molecular Cloud.  They were able to identify 7
embedded clusters associated with the centers of 7 massive clumps of
molecular gas identified previously by Williams, Blitz \& Stark
(1995)~\cite{ref:williams95}.  These 7 clusters were all associated
with far IR IRAS sources, and 5 were previously unknown. Thus what
appeared to be single point sources in the IRAS data turned out to be
embedded star clusters. Figure \ref{fig:ros.ps} shows a plot of the
clumps in the Rosette vs. the gravitational boundedness of the clumps,
plotted as $M_{grav}/M_{lum}$ where $M_{grav} = RV^2 / G,~ M_{lum} = 
X\int\int\int T_A dv dx dy$ and $X$ is the usual CO-to-H$_2$ conversion
factor.

The clusters identified by Phelps \& Lada~\cite{ref:phelps97} are
identified with the most massive, gravitationally bound clumps
(Williams et al., 1995)~\cite{ref:williams95}.  But what is it about
the star-forming clumps that produces a great many star-forming cores
simultaneously?  In other words, what is it that is communicated
through a clump in a crossing time to let all parts know that they
must produce stars simultaneously?  What determines how many stars
form within a given clump? Do the clumps even have embedded cores that
are distinct, recognizable entities?  The Phelps \& Lada work also
provides an efficient way to find embedded clusters, and,  {\it en
passant}, demonstrates that the clumps are real, long-lived entities,
rather than ephemeral turbulent structures, as some authors have
suggested (otherwise the star clusters would not have had enough time
to form in them).

\subsection{Star Formation on Galactic Scales}

Understanding clustered star formation will likely solve the problem
of how star formation takes place within an individual molecular
cloud.  How then do we extrapolate to larger scales, to the scale of an
entire galaxy?  A reasonable question to ask is whether we need to
know all of the details of the star formation process to address star
formation on galactic scales.  That is,
since we know that star formation takes place only in molecular
clouds, and that the star formation efficiency in molecular clouds
tends to be small ($\sim 5\%$), with relatively little variation 
in normal galaxies, perhaps
the question of how stars form in galaxies reduces to a question of how the
molecular clouds themselves form?  That is, if we can understand how
the ISM turns molecular gas into GMCs, and we can understand how the
different conditions within GMCs translate into different star
formation efficiencies and perhaps even IMFs, then it should be
possible to determine the global star formation rate from just the gas
content and other physical conditions within the galaxies.

\begin{figure}[!htb]
\begin{center}
\vskip 1.0cm
\psfig{figure=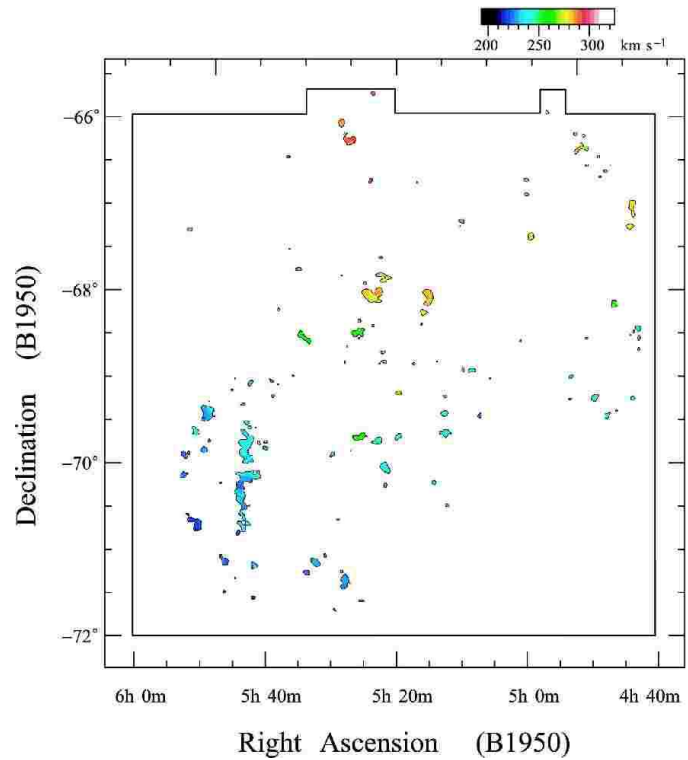,height=10cm}
\caption{Map of the GMCs found in the LMC by Mizuno et al. 
(2001).  These cover essentially the entire face
of the galaxy, but cover only a small fraction of the surface,
necessitating a large effort to obtain the map.
\label{fig:lmc}}
\end{center}
\end{figure}

\begin{figure}[!htb]
\begin{center}
\vskip 1.0cm
\psfig{figure=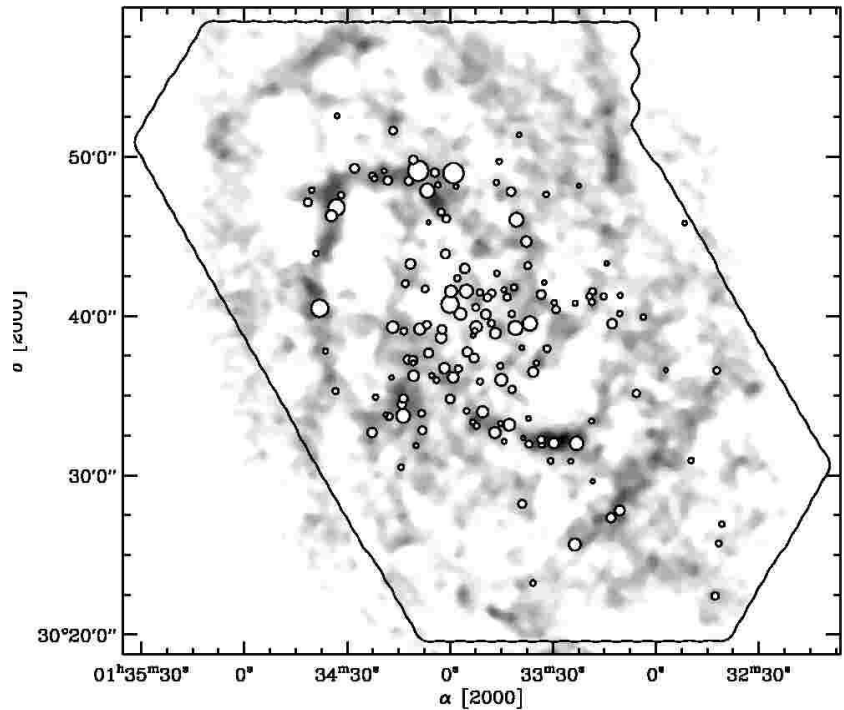,height=10cm}
\caption{The molecular clouds catalogued by Engargiola et al.~(2003),
shown as white dots enclosed by black circles, superimposed on an
HI map from the data of Deul and van der Hulst (1987).  The
area  of each dot is proportional to the H$_2$ mass of each GMC.
Notice both the filamentary structure of the HI and 
the good correspondence between the filaments and the
location of the GMCs.
\label{fig:m33h1}}
\end{center}
\end{figure}

To this end, it is useful to have complete surveys of individual GMCs
in entire galaxies not just unresolved images of the molecular gas,
This has become possible only in the last few years,  but for only a
few galaxies; only two such maps have been published.  The first was
the LMC which has been nearly completely mapped by Mizuno et al.
(2001)~\cite{ref:mizuno01} using the 4m Nanten telescope (see Figure
\ref{fig:lmc}).  More recently, Engargiola et al.
(2003)~\cite{ref:engargiola03} have used the BIMA array to make a 759
field mosaic of M33 at 15\asece resolution ($\sim 50$~pc -- see Figure
\ref{fig:m33h1}).  Both of these images indicate the difficulty in
surveying galaxies for individual GMCs:  the surface filling fraction
of GMCs in galactic disks is small (see Figure \ref{fig:lmc}), and the
resolution needed to determine the cloud properties is high, requiring
either large amounts of telescope time for Local Group objects, or
high sensitivity interferometric mosaics for galaxies farther away.
For both the LMC and M33, followup observations at high resolution
were needed to resolve the molecular clouds in each case.

Other galaxies that have been fully mapped to date but not yet published
include the SMC (Mizuno et al.), IC10 (Leroy et al.), and M31
(Muller and Guelin);  these maps were all presented at this YLU conference.
Although there has been a herculean effort to
map the molecular gas in M31, the resolution (90 pc) appears to be too
low to resolve clouds blended in the beam in many directions (Muller,
this conference).  Followup interferometric observations will be
needed to obtain the properties of the GMCs.
The central region of M64 has also been mapped at 
high enough resolution to 
measure the molecular cloud properties in the nuclear region where
the surface filling fraction of molecular gas approaches unity 
(Rosolowsky \& Blitz 2005)~\cite{ref:rosolowsky05}.  The disk has 
not been observed at comparably high resolution.

The image of M33 seen in Figure \ref{fig:m33h1} shows something quite 
striking and new:  essentially all of the individual GMCs lie on 
filaments of HI.  Note, though, that
the filaments show little variation in surface density with radius, 
but that the GMCs become very sparse 
at radii more than about 12\amine from the
center.  Averaged over annuli, the atomic gas surface density is nearly
constant with radius, falling by only a factor of two over 7 kpc,
but the molecular gas surface density is exponential with a scale 
length of 1.4 kpc.  Because there is a
great deal of HI where there is no CO, the H$_2$ must have formed from
the HI, rather than the converse.  But why do the GMCs become so
sparse beyond about 3 kpc?

The close association of the molecular clouds with the filaments
implies a maximum lifetime for the GMCs of $\sim$ 20 Myr, based on the
mean velocity difference between the CO and HI along the same line of
sight.  A significantly longer lifetime would cause a spatial
separation between the atomic and molecular gas.    
It thus appears that the filaments
are a necessary, but not  sufficient condition for the formation of
molecular clouds.  What, produces the radial abundance gradient 
of molecular gas, and thus the radial variation of the star formation 
rate?  

One possibility is that the filaments are really the boundaries of 
`holes', large regions relatively devoid of HI, caused by
supernova explosions in a previous generation of OB associations.
However, the large holes in 
Figure \ref{fig:m33h1} are not associated with catalogued OB associations
(Deul \& van der Hulst 1987)~\cite{ref:deul87}.  
In any event, energies of $\sim$10$^{53}$ ergs are needed to evacuate the 
large holes, implying that 100 or more O stars
would have been formed in each, leaving bright stellar
clusters and diffuse x-ray emission at the centers of the emply regions, 
which are not observed.

\begin{figure}[!htb]
\begin{center}
\vskip 1.0cm
\psfig{figure=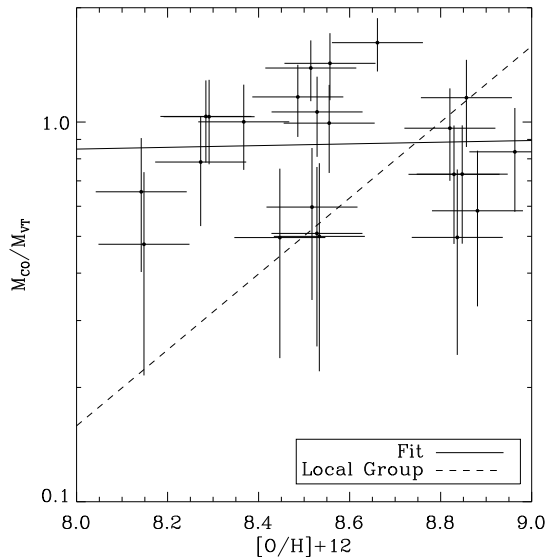,height=8cm}
\caption{A plot of the X-factor as a function of metallicity in M33.
The X-factor is given as a ratio of the value in M33 vs. the locally
determined value of 2 $\times 10^{20}$ cm$^{-2}$ (K \kms)$^{-1}$. 
The dashed line plots the trend from Arimoto et al. (1996) summarizing
similar measurements throughout the Local Group.
\label{fig:m33x}}
\end{center}
\end{figure}

Could it be that the radial variation is due to a change in the
ratio of CO/H$_2$, the so-called ``X'' factor produced by the known
abundance gradient in M33?  This possibility was investigated
by Rosolowsky et al. (2003)~\cite{ref:rosolowsky03} who showed that if $X$ is determined
by equating the luminous CO mass with the virial mass of resolved
clouds in M33, $X$ shows no variation with metallicity or radius.
This can be seen from Figure \ref{fig:m33x}.

Wong \& Blitz (2002)~\cite{ref:wong02} have proposed that the fraction of
molecular gas at a particular radius in a galaxy is the result of
interstellar pressure, based on interferometric observations of six
nearby spiral galaxies.  Blitz \& Rosolowsky (2004)~\cite{ref:blitz04} 
showed that pressure modulated molecular cloud formation
implies that the radius in a galaxy where the atomic/molecular surface
density is unity should occur at a constant {\it stellar} surface 
density.  An investigation of 30 galaxies showed this constancy to be
good to within 50\%.  Thus it seems reasonable to conclude that
hydrostatic pressure plays a significant role in the formation of molecular
clouds. 

\begin{figure}[!htb]
\begin{center}
\vskip 1.0cm
\psfig{figure=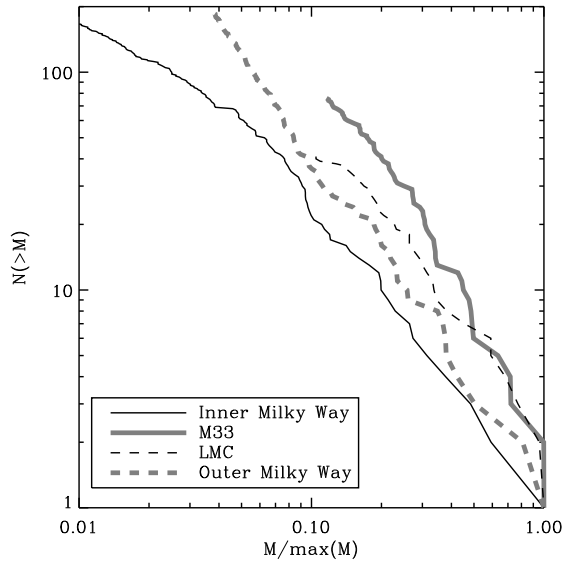,height=8cm}
\caption{The cumulative mass distribution in three Local Group
galaxies, with the inner and outer Milky Way plotted separately.
This plot shows that there are significant differences in the mass
spectrum from galaxy to galaxy, with the inner Milky Way giving a
power law in dN/dM of -1.6, and M33 giving a power law index of -2.3.
\label{fig:mspec}}
\end{center}
\end{figure}

But if hydrostatic pressure is the main culprit in forming GMCs,
how do the GMCs vary from galaxy to galaxy where interstellar pressure
might be quite varied?   With current telescopes,
we have data only for GMCs in the Local
Group galaxies, and the published data are only available for the Milky Way,
the LMC and M33.  If we examine the cumulative mass 
distribution of GMCs for each galaxy (but
separating the inner Milky Way from the outer Milky Way), we see that
there are significant differences 
from one galaxy to another (Figure \ref{fig:mspec}).  In this
figure, the mass distribution is normalized to the most massive cloud
observed, and the distribution for M33 is significantly steeper
than that of the other galaxies.  The mass function
is independent of resolution, and the differences in slope are significant.  

\begin{figure}[!htb]
\begin{center}
\vskip 1.0cmr
\psfig{figure=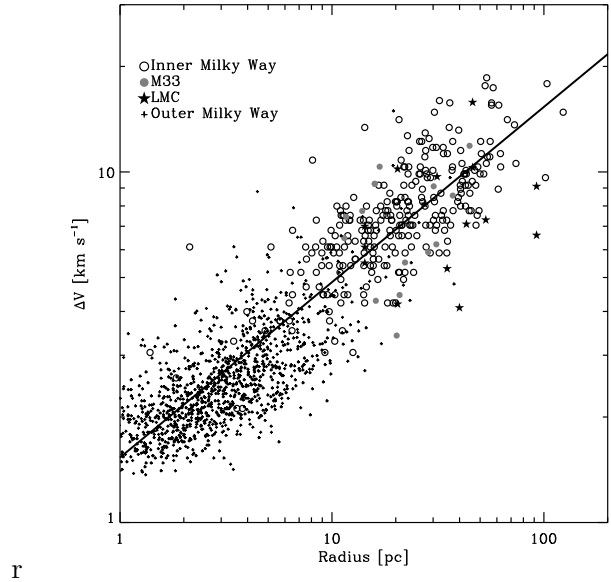,height=8cm}
\caption{Size-linewidth relation for resolved GMCs in the same
galaxies plotted in Figure \ref{fig:mspec}.  The clouds in all three
galaxies appear to follow the same relation with $\Delta V \propto 
R^{1/2}$. Corrections have been applied to the published data to treat
all of the data identically, and to correct for beam effects.
\label{fig:rdv}}
\end{center}
\end{figure}

We may also ask whether the clouds show differences, for
example, in the size-linewidth relation observed for clouds in the
Milky Way.  Figure \ref{fig:rdv} shows a plot of hundreds of clouds
in the Milky Way, M33, and the LMC, with a line of slope 1/2
superimposed on the data.  Evidently, the clouds in these galaxies
obey the same size-linewidth relation with no zero-point offset:
$\Delta V \propto R^{1/2}$.  

This plot suggests that if all of the clouds in these galaxies are
self-gravitating, the surface density of the clouds is
constant with a relative scatter given by the scatter in 
Figure \ref{fig:rdv}.  That is, since $\Delta V \propto R^{1/2}$, and  
$M \propto R(\Delta V)^2$/G, then $M/R^2 = const$. But the mean
internal pressure of GMCs can be written: $P_{int}$ = $\alpha (\pi/2)  
G {\Sigma_g}^2$, where $\Sigma_g$ is the gas surface density of the
clouds and $\alpha$ is a constant near unity that depends on the cloud
geometry.  Thus, the GMCs that compose Figure \ref{fig:rdv} have the
same mean internal pressure, regardless of size, regardless of the
galaxy they are in and regarless of the external pressure.

This gives us a way of understanding how the IMF might indeed be
constant from galaxy to galaxy, at least for galaxies similar to those
in Figure \ref{fig:rdv}. That is, if the mean internal pressure of all
GMCs in the disk of a galaxy is the same, then the range of pressures
within a GMC might also be the same, and the star-forming cores might
therefore also be quite similar.  It is important to keep in mind,
however, that even if true it might apply only in the disks of
galaxies.  In the bulge regions, the hydrostatic pressure of the gas
is likely to be two to three orders of magnitude higher than that in
the disk (e.g.  Spergel \& Blitz (1992)~\cite{ref:spergel92}.  In
these regions, the external pressure can significantly exceed the mean
internal pressure of a few $\times 10^5$ cm$^{-3}$ K of the clouds in
the disk.  In the bulge regions, the GMCs must be different from those
in the disk, and may well give rise to stars with a different IMF.

Studying global star formation is only in its infancy and new
instruments coming on line and being developed should provide
the sensitive high resolution data needed to get from single star
formation to the Madau plot.  Equally important is to have those
who work on local star formation interact closely with those working
on global star formation on a regular basis as has happened in this
conference.

\section*{Acknowledgments}
I'd like to thank Charlie Lada who commented on an early version of
this manuscript and Steve Stahler for a useful discussion.
Erik Rosolowsky prepared a number of the figures in this paper.

\end{document}